\documentclass[aps,prb,superscriptaddress,amsfonts,amsmath,amssymb,floatfix,reprint,longbibliography]{revtex4-2}

\usepackage{url}
\usepackage{amsmath}
\usepackage{amstext}
\usepackage{amssymb}
\usepackage{amsfonts}
\usepackage{amsbsy}
\usepackage{verbatim}
\usepackage{color}
\usepackage[colorlinks=true, urlcolor=blue, linkcolor=blue, citecolor=blue, pdftex]{hyperref}
\usepackage{floatrow}
\usepackage{float}
\usepackage{tikz}
\usepackage{gensymb}
\usepackage{textcomp}
\usepackage{enumitem}
\usepackage[version=3]{mhchem}
\usepackage{afterpage}
\usepackage{siunitx}
\usepackage{latexsym}
\usepackage[misc]{ifsym}
\usepackage[normalem]{ulem}
\usepackage{times}
\usepackage{braket}
\usepackage{tikz}
\usepackage{tikz-3dplot}

\newcommand{\siteOne}{
	\begin{tikzpicture}  
		\draw[] ( 60 : 0.125) to (240 : 0.125);
		\draw[] (120 : 0.125) to (300 : 0.125);
		\draw[thick, fill = black] (0, 0) circle (0.03);
	\end{tikzpicture}
}
\newcommand{\squareSites}{
	\begin{tikzpicture}
      \begin{scope}[shift = {(0, 1.50)}]
		\draw[] (-0.125, -0.125) to ($(-0.125, -0.125) + (150 : 0.125)$);
		\draw[] (-0.125, -0.125) to ($(-0.125, -0.125) + (300 : 0.125)$);
		\draw[] (-0.125, +0.125) to ($(-0.125, +0.125) + (210 : 0.125)$);
		\draw[] (-0.125, +0.125) to ($(-0.125, +0.125) + ( 60 : 0.125)$);
		\draw[] (+0.125, -0.125) to ($(+0.125, -0.125) + (240 : 0.125)$);
		\draw[] (+0.125, -0.125) to ($(+0.125, -0.125) + ( 30 : 0.125)$);
		\draw[] (+0.125, +0.125) to ($(+0.125, +0.125) + (330 : 0.125)$);
		\draw[] (+0.125, +0.125) to ($(+0.125, +0.125) + (120 : 0.125)$);
		\draw[] (-0.125, -0.125) to (-0.125, +0.125) to (+0.125, +0.125) to (+0.125, -0.125) to (-0.125, -0.125);
		\draw[thick, fill = black] (-0.125, -0.125) circle (0.03);
		\draw[thick, fill = black] (-0.125, +0.125) circle (0.03);
		\draw[thick, fill = black] (+0.125, -0.125) circle (0.03);
		\draw[thick, fill = black] (+0.125, +0.125) circle (0.03);
      \end{scope}
	\end{tikzpicture}
}
\newcommand{\siteSix}{
	\begin{tikzpicture}
		\draw[] (150 : 0.125) to (330 : 0.125);
		\draw[] (210 : 0.125) to ( 30 : 0.125);
		\draw[thick, fill = black] (0, 0) circle (0.03);
	\end{tikzpicture}
}

\begin{document}

\title{Tensor network study of the spin-\texorpdfstring{$1/2$}{TEXT} Heisenberg anti-ferromagnet on the Shuriken lattice}

\author{Philipp Schmoll}
\affiliation{Dahlem Center for Complex Quantum Systems and Institut f\"{u}r Theoretische Physik, Freie Universität Berlin, Arnimallee 14, 14195 Berlin, Germany}

\author{Augustine Kshetrimayum}
\affiliation{Dahlem Center for Complex Quantum Systems and Institut f\"{u}r Theoretische Physik, Freie Universität Berlin, Arnimallee 14, 14195 Berlin, Germany}
\affiliation{Helmholtz-Zentrum Berlin f\"{u}r Materialien und Energie, Hahn-Meitner-Platz 1, 14109 Berlin, Germany}
\affiliation{Theory Division, Saha Institute of Nuclear Physics, 1/AF Bidhannagar, Kolkata 700 064, India}

\author{Jan Naumann}
\affiliation{Dahlem Center for Complex Quantum Systems and Institut f\"{u}r Theoretische Physik, Freie Universität Berlin, Arnimallee 14, 14195 Berlin, Germany}

\author{Jens Eisert}
\affiliation{Dahlem Center for Complex Quantum Systems and Institut f\"{u}r Theoretische Physik, Freie Universität Berlin, Arnimallee 14, 14195 Berlin, Germany}
\affiliation{Helmholtz-Zentrum Berlin f\"{u}r Materialien und Energie, Hahn-Meitner-Platz 1, 14109 Berlin, Germany}

\author{Yasir Iqbal} 
\email[]{yiqbal@physics.iitm.ac.in}
\affiliation{Department of Physics and Quantum Centers in Diamond and Emerging Materials (QuCenDiEM) group, Indian Institute of Technology Madras, Chennai 600036, India}

\date{\today}

\begin{abstract}

We investigate the ground state of the spin $S=1/2$ Heisenberg anti-ferromagnet on the Shuriken lattice, also in the presence of an external magnetic field. To this end, we employ two-dimensional tensor network techniques based on infinite projected entangled pair and simplex states considering states with different sizes of the unit cells. We show that a valence bond crystal with resonances over length six loops emerges as the ground state (at any given finite bond dimension) yielding the lowest reported estimate of the ground state energy $E_0/J = -0.4410 \pm 0.0001$ for this model, estimated in the thermodynamic limit. We also study the model in the presence of an external magnetic field and find the emergence of $0$, $1/3$ and $2/3$ magnetization plateaus with states respecting translation and point group symmetries that feature loop-four plaquette resonances instead.

\end{abstract}

\maketitle
 
\section{Introduction}
Systems of anti-ferromagnetically interacting quantum spins decorated on corner sharing arrangements of triangles continue to attract much interest as promising platforms for realizing novel quantum phases~\cite{Savary_2016}. Indeed, the arrival of candidate quantum spin liquid materials based on the iconic 
Kagome lattice such as the celebrated Herbertsmithite~\cite{Mendels-2007,Han-2012,Khuntia-2020} and Kapellasite~\cite{Fak-2012} have provided an impetus to the field of frustrated magnetism. Their intriguing properties have 
triggered a flurry of experimental and theoretical studies which established the Kagome lattice as a fertile host for a myriad of exotic states. The parameter space of its Heisenberg Hamiltonian in the presence of long-range interactions is known to be host to quantum spin liquids including chiral states, spin and lattice nematics, and valence bond crystals. Recently, a class of materials based on a different corner sharing arrangement of triangles \textemdash\ the so called \emph{Shuriken lattice} (also called square-Kagome, Squagome, and squa-Kagome lattice.) \textemdash\ have come into limelight as promising candidate \emph{quantum spin liquid materials}~\cite{Yakubovich-2021,PhysRevB.94.014429}. No sign of magnetic ordering down to 50 mK has been observed in the spin $S=1/2$ Cu$^{2+}$ based materials KCu$_6$AlBiO$_4$(SO$_4$)$_5$Cl~\cite{Fujihala-2020} and Na$_6$Cu$_7$BiO$_4$(PO$_4$)$_4$[Cl,(OH)]$_3$~\cite{Liu-2022} despite them having large negative Curie-Weiss temperatures of $\SI{-237}{\kelvin}$ and $\SI{-212}{\kelvin}$, respectively. This reveals a scenario similar to Herbertsmithite for which dominant anti-ferromagnetic interactions on this highly frustrated lattice prevent the onset of magnetic order. Such studies can be traced back to early work hinting at quantum materials featuring that lattice structure
\cite{PhysRevB.5.4472}.
On the theoretical front, previous investigations into the nature of the ground state of the $S=1/2$ Heisenberg anti-ferromagnet have provided compelling evidence for a magnetically disordered ground state while revealing a subtle competition between different types of nonmagnetic ground states which remains debated~\cite{Astrakhantsev2021,Lugan-2019,Morita-2018,Hasegawa-2018,Ralko-2015,Rousochatzakis-2013,Nakano-2013,Richter-2009,Tomczak-2003}. \\

In this work, we employ instances of two-dimensional \emph{tensor network} (TN) algorithms formulated directly in the thermodynamic limit towards resolving the nature of the ground state. Tensor networks are quantum-information inspired tools that use entanglement as a resource for studying strongly correlated quantum many-body systems~\cite{Orus-AnnPhys-2014,EisertTensorNetworks,AreaReview,VerstraeteBig}. They naturally build in quantum correlations and are suited to capture non-local entanglement by construction. They do not suffer from the sign problem plaguing quantum Monte Carlo simulations on frustrated systems. Moreover, these techniques can be used to study large system sizes including the thermodynamic limit, thus mitigating finite size effects. In two spatial dimensions, they are known as \textit{projected entangled pair states} (PEPS) or iPEPS~\cite{Jordan2008,Verstraete-arxiv-2004} in their infinite instance and have become a state-of-the-art numerical tool for studying strongly interacting systems. These techniques have recently proven to be quite successful in studying frustrated model Hamiltonians~\cite{Xiangkhaf,ThibautspinS,Thibautnematic,KshetrimayumkagoXXZ,Schmollsu2benchmark}, real materials~\cite{CorbozMaterial,Kshetrimayummaterial,SchmollfinT2022}, open systems~\cite{Kshetrimayum-2017,piotr2012,piotr2015,piotr2016,Kshetrimayumthermal,MondalPRB2020} and non-equilibrium phenomena~\cite{Czarnikevolution2019,Hubig2019,Kshetrimayum2DMBL,Kshetrimayum2DTC,Dziarmaga2021,Dziarmagaevolv2022}.

\begin{figure*}[t]
    \begin{minipage}{0.3\textwidth}
        \includegraphics[width = 0.9\columnwidth]{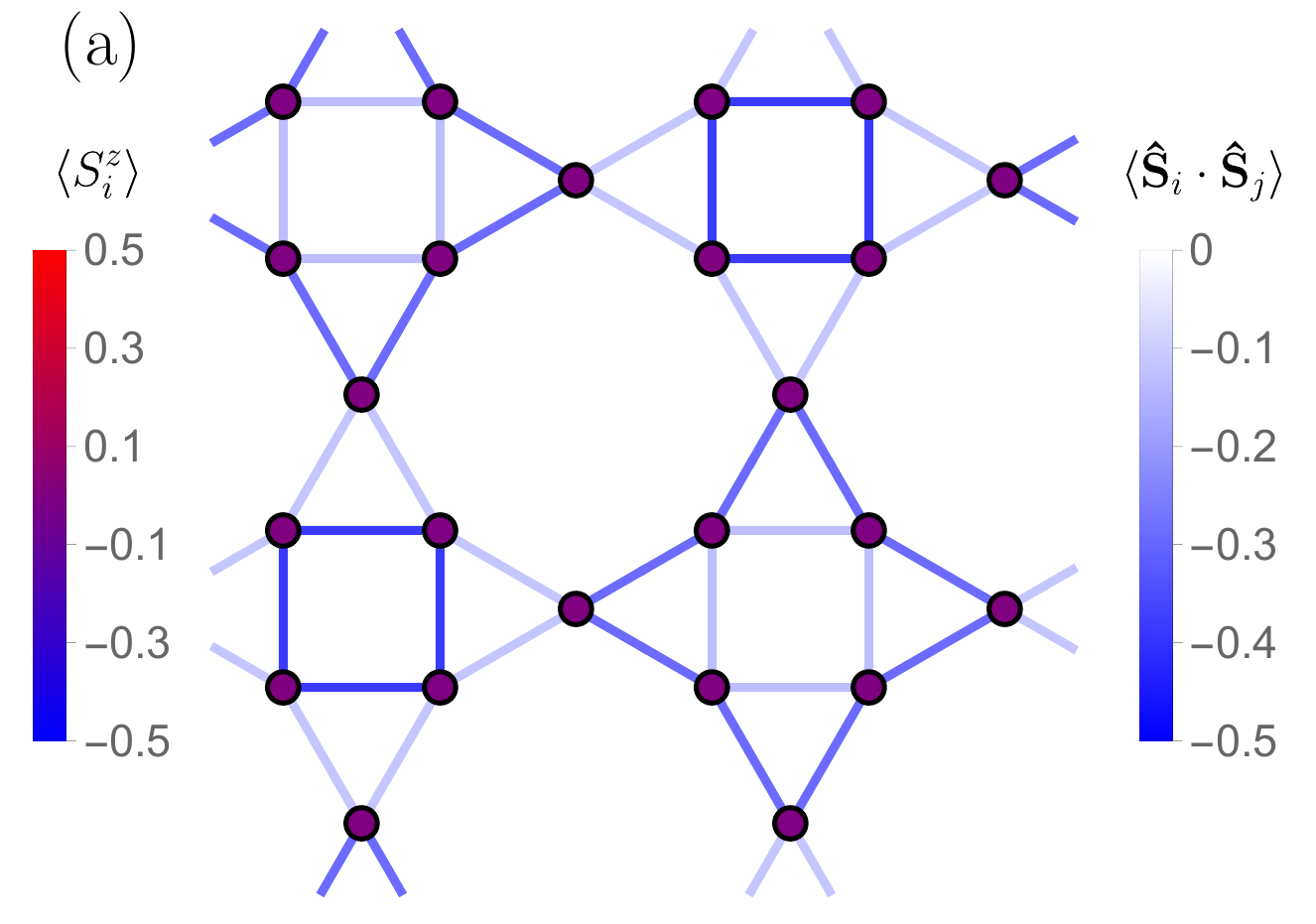}
    \end{minipage}
    \begin{minipage}{0.3\textwidth}
        \includegraphics[width = 0.9\columnwidth]{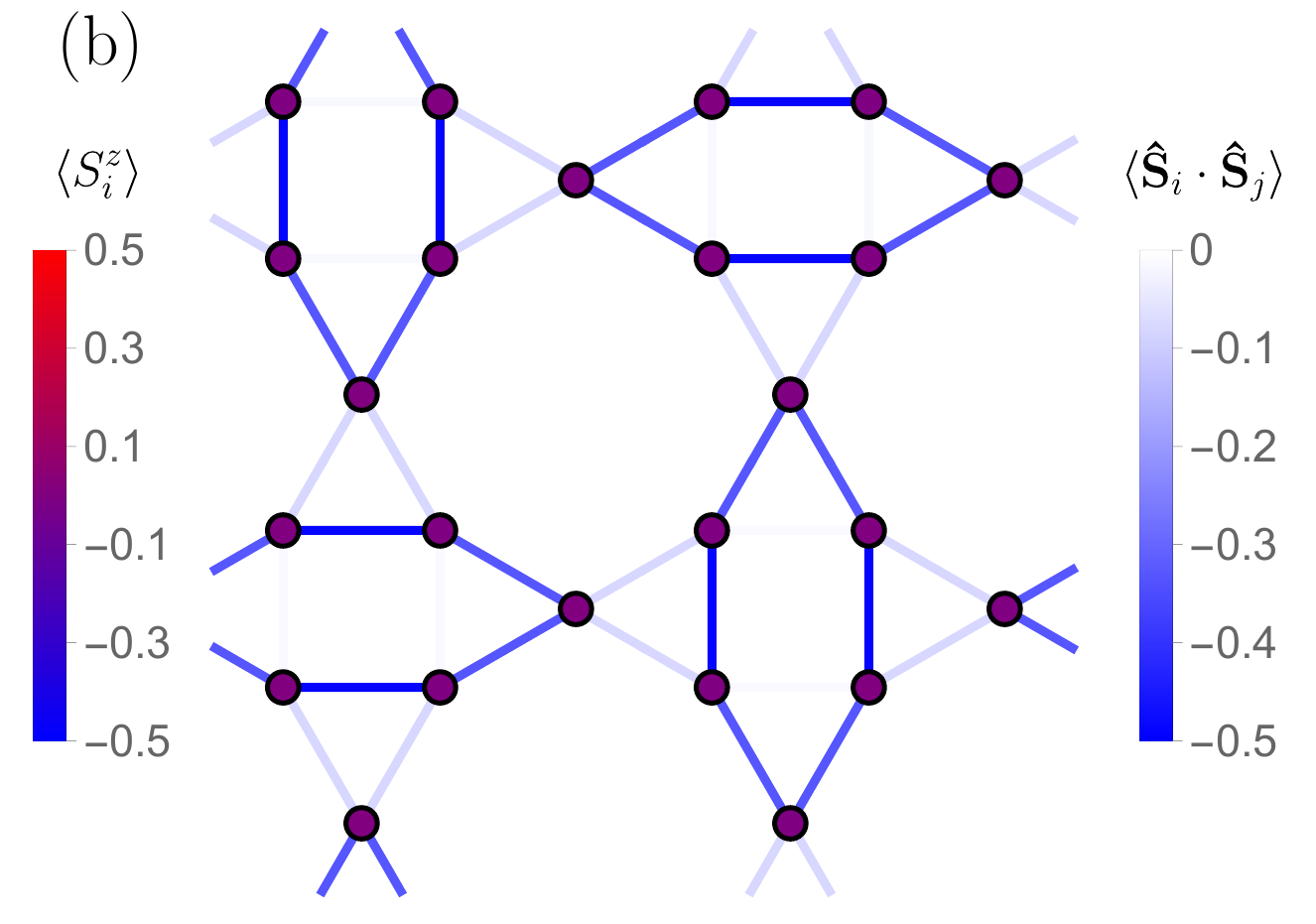}
    \end{minipage}
    \begin{minipage}{0.38\textwidth}
        \includegraphics[width = 0.92\columnwidth]{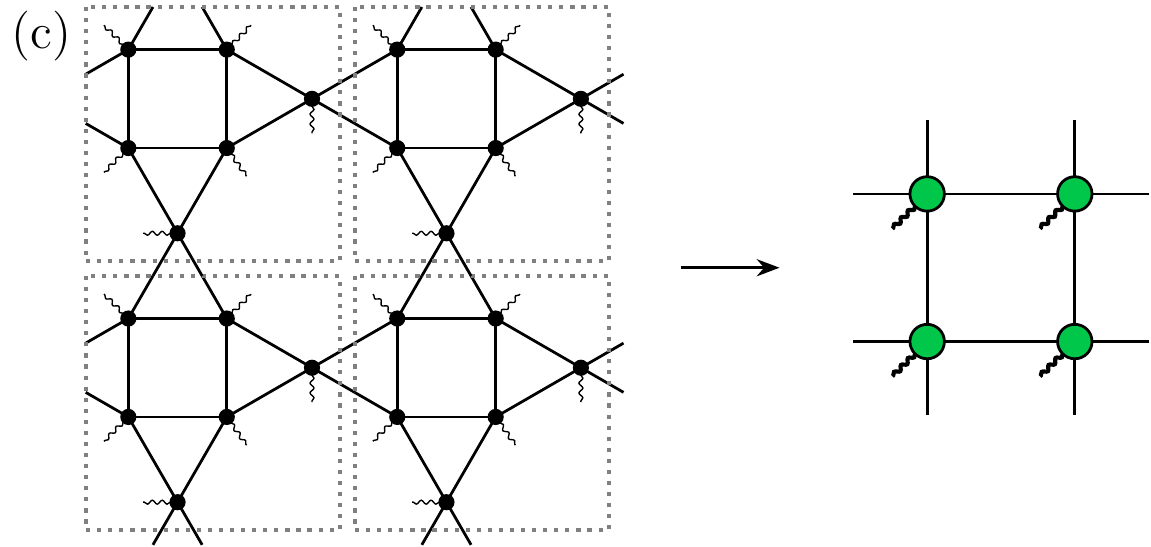}
    \end{minipage}
    \caption{Nearest-neighbour spin-spin correlations for the ground state configuration of the pinwheel VBC (a) and the loop-six VBC (b) states at $\chi_B = 12$. The expectation values of the spin-$z$ component are typically $< 10^{-3}$ and only shown for completeness. (c) Shuriken lattice with spin-$1/2$ degrees of freedom on the sites. The elementary unit cell (gray rectangles) consists of six sites, which are coarse-grained to map the 
    Shuriken lattice to a regular square lattice. Straight lines denote virtual bond indices, curly lines denote physical indices in the TN structure.}
    \label{fig:groundStateConfigurations_1}
\end{figure*}

Recent theoretical works investigating the ground state of the isotropic $S=1/2$ Heisenberg anti-ferromagnet on the Shuriken lattice have identified two competing \emph{valence bond crystals} (VBCs) involving resonating loops of different lengths: (i) a pinwheel VBC which maximizes the number of smallest possible loops of length four [see Fig.~\ref{fig:groundStateConfigurations_1}(a)] and (ii) a VBC pattern comprising {\it only} of loops of length six [see Fig.~\ref{fig:groundStateConfigurations_1}(b)]. Surprisingly, it has been shown within an effective \emph{resonating valence bond} (RVB) theory that the tunneling processes can be renormalized in such a way that the smallest loops are not always the most relevant in capturing the correct ground state correlations~\cite{Ralko-2015}. Indeed, based on a quantum dimer model approach it has been shown in Ref.~\cite{Ralko-2015} that the loop-six VBC is more stable energetically compared to the pinwheel VBC when non-local processes outside the nearest-neighbor valence bond basis were 
invoked. Complementing this, based on energetic considerations alone, the pinwheel VBC should conventionally be the expected ground state as found in a recent variational Monte Carlo study~\cite{Astrakhantsev2021}. This opens a delicate question on how to properly account for such non-local quantum correlations and patterns of long-range entanglement in highly degenerate frustrated systems.

Here, we use a \emph{tensor network} approach to simulate the model directly in the thermodynamic limit. TNs represent the state vector of a many-body system, e.g., reflecting the ground state, as a contraction of a network of local tensors, that are connected by auxiliary indices (bond indices). This enables efficient numerical simulations with only a polynomial scaling in the number of constituents~\cite{VerstraeteBig,RevModPhys.93.045003,Orus-AnnPhys-2014,EisertTensorNetworks}. In this work, we employ the 
\emph{infinite projected entangled pair state} (iPEPS)~\cite{Jordan2008} and 
\emph{infinite projected entangled simplex state} (iPESS) (a variant of iPEPS)~\cite{Xiangpess} techniques with an ansatz based on different and specifically tailored unit cell sizes for optimizing the ground state of our model. In this context, the TN is used as an ansatz for the full many-body state vector, consisting of a unit cell of different tensors that generates a translationally invariant state. The accuracy of the ansatz can be systematically improved by increasing the \textit{bond dimension} of the TN, which is the dimension of the virtual indices connecting the local tensors, see Fig.~\ref{fig:groundStateConfigurations_1}(c) [see Appendix]. It controls the number of variational parameters in the ansatz and is a measure for the amount of quantum entanglement that can be captured. {We mainly employ the so-called simple update~\cite{simpleupdatejiang} to optimize the ground state tensors, which is expected to work well for the gapped model at hand~\cite{CorbozFermions2010,Weichselbaum,simpleupdatejiang}. In order to verify its accurate functioning and ability to resolve the close competition between the two candidate ground states, we additionally employ a variational update~\cite{Naumann2023} for this task.} 
The \textit{corner transfer matrix renormalization group} (CTMRG)~\cite{ctmnishino1996,ctmnishino1997,ctmroman2009,ctmroman2012} is then used to compute the expectation values of the ground state energy in a variational manner, the spin-$z$ operator as well as the two-point correlations to decipher the nature of the ground state. We also employ additional $SU(2)$-symmetric simulations~\cite{Schmoll2020,Schmollsu2benchmark} for the model. Given the flexibility of the framework, we apply an external magnetic field to study the magnetization process of the model and provide a compelling picture of the nature of phases corresponding to different magnetization plateaus. 

\section{Model and methods}

The model we are considering is the $S=1/2$ Heisenberg anti-ferromagnet on the Shuriken lattice 
\begin{equation}
    \hat{H} = \sum_{\langle i,j \rangle} \mathbf{\hat S}_i \cdot \mathbf{\hat S}_j - h \sum_i {\hat S}^z_i\  \label{eq:heisenberg}
\end{equation}
in the presence of an external magnetic field, where $\mathbf{\hat S}_i$ are the $S=1/2$ operators on site $i$ and $\langle i,j \rangle$ denotes nearest-neighbours. The Shuriken lattice [see Fig.~\ref{fig:groundStateConfigurations_1}(c)] features corner-sharing triangles, and thus leads to only a marginal alleviation of geometric frustration in the presence of anti-ferromagnetic couplings. Being composed of corner-sharing triangles, it is locally similar to the Kagome lattice. At the same time, the Shuriken lattice shares two inequivalent sublattices, rendering this lattice ideal to study effects of lattice anisotropy, for which our methods are ideally suited.

We have applied two different TN structures for the simulation of the Shuriken lattice. The first ansatz (iPEPS) uses a partial coarse-graining of the Shuriken lattice to an irregular square lattice. Inspired by the success for the \mbox{$S=1/2$} Kagome Heisenberg anti-ferromagnet~\cite{Xiangkhaf}, the second structure is based on the iPESS ansatz~\cite{Xiangpess} that generalizes iPEPS to lattices with higher simplices. For the Shuriken lattice, it is defined on its dual lattice, the so-called $(4, 8^2)$ Archimedean lattice (also referred to as the square-octagon, Fisher or CaVO lattice). While the simple update for iPESS incorporates three lattice sites at each update step, it includes six sites for iPEPS [see Appendix]. In order to compute expectation values, a unit cell of six sites on the Shuriken lattice is coarse-grained into a single tensor on the regular square lattice, as shown in Fig.~\ref{fig:groundStateConfigurations_1}(c). This approach is taken both for the iPESS and the iPEPS simulations, starting from a $(4, 8^2)$ Archimedean lattice and a partially coarse-grained Shuriken lattice, respectively. A directional CTMRG routine then computes effective environment tensors for each coarse-grained iPEPS tensors, such that quantum correlations are fully incorporated when computing expectation values (details are given in the Appendix). This is achieved by a well chosen environment bond dimension $\chi_E$ \textemdash\ a refinement parameter controlling the approximations in the CTMRG routine, which is increased until the expectation values converge.

\section{Results on the ground state energy and dimer orders}

The 
ground state energy of the Shuriken Heisenberg model can be straightforwardly evaluated in the TN representation. The Hamiltonian consists of a sum of nearest-neighbour terms,
\begin{align}
    E_0 = \frac{1}{N} \sum_{\langle i, j \rangle} \langle \psi_0 \vert h_{i,j} \vert \psi_0 \rangle\,
\end{align}
where $N$ is the number of lattice sites and $\vert \psi_0 \rangle$ is the normalized ground state vector. The smallest possible geometrical unit cell of the Shuriken lattice consists of six sites [the dashed regions in Fig.~\ref{fig:groundStateConfigurations_1}(c)]. This ansatz --- which 
imposes translational invariance while being compatible with quantum spin liquid and lattice nematic candidate ground 
states --- would, however, fail to capture translation symmetry broken VBC orders such as the pinwheel and the loop-six VBCs, which are the prime competing ground state candidates. Therefore, we use different unit cell sizes to search for competitive TN ansätze with the lowest ground state energy. The different unit cell configurations are then labeled by the size of the super-unit-cell on the square lattice, denoted by $(L_x, L_y)$. A configuration $(L_x, L_y)$ hence corresponds to an TN state with $6 \cdot L_x \cdot L_y$ spins in total.

\begin{figure}[b]
    \includegraphics[width = 1.0\columnwidth]{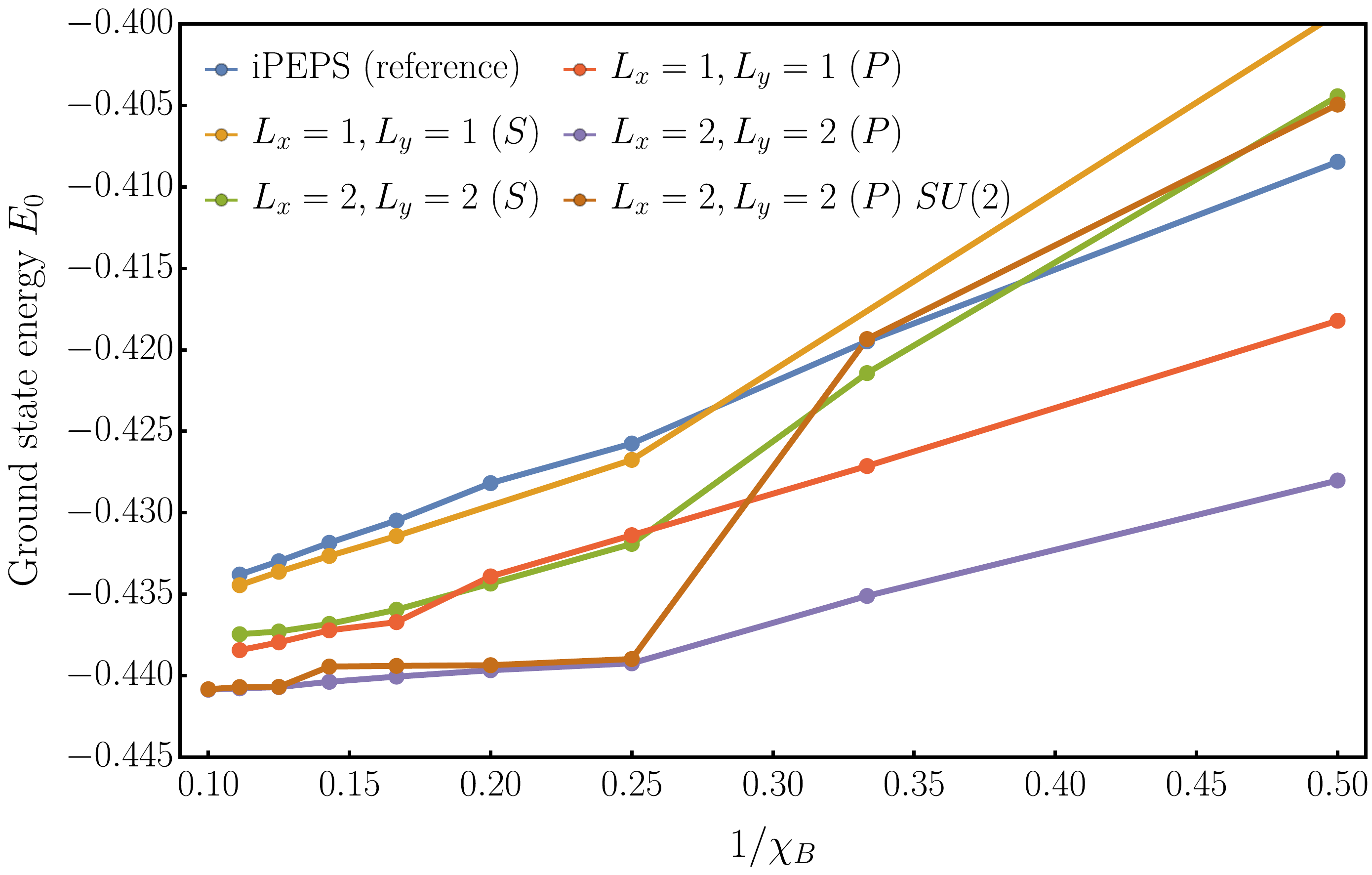}
    \caption{Ground state energy without magnetic field [$h=0$ in Eq.~\eqref{eq:heisenberg}] versus the inverse of the bond dimension $\chi_B$. 
    iPESS simulations are denoted by $(S)$, iPEPS simulations by $(P)$.}
    \label{fig:groundStateEnergy_1}
\end{figure}

In Fig.~\ref{fig:groundStateEnergy_1}, we show the ground state energy for the iPEPS and iPESS simulations and square lattice unit cells of $(1, 1)$ and $(2, 2)$ as a function of the inverse of the iPEPS bond dimension $\chi_B$ (bulk bond dimension). Our results are compared to a previous iPEPS study of the model in Ref.~\cite{Astrakhantsev2021}, using a $(1, 1)$ TN ansatz with a coarse-graining to a honeycomb lattice. Based on our simulations with different sizes of the unit cells, we find the lowest energy is obtained with a $(2,2)$ configuration (unit cell with 24 sites), i.e., corresponding to a valence bond crystal ground state. 
A similar ground state can be obtained by a $(1, 2)$ configuration in a checkerboard arrangement, which consists of only twelve spins. However, we use the more general state vector ansatz with 24 spins to be able to incorporate possible richer patterns of spin correlations.

\begin{figure}[t]
    \includegraphics[width = 1.0 \columnwidth]{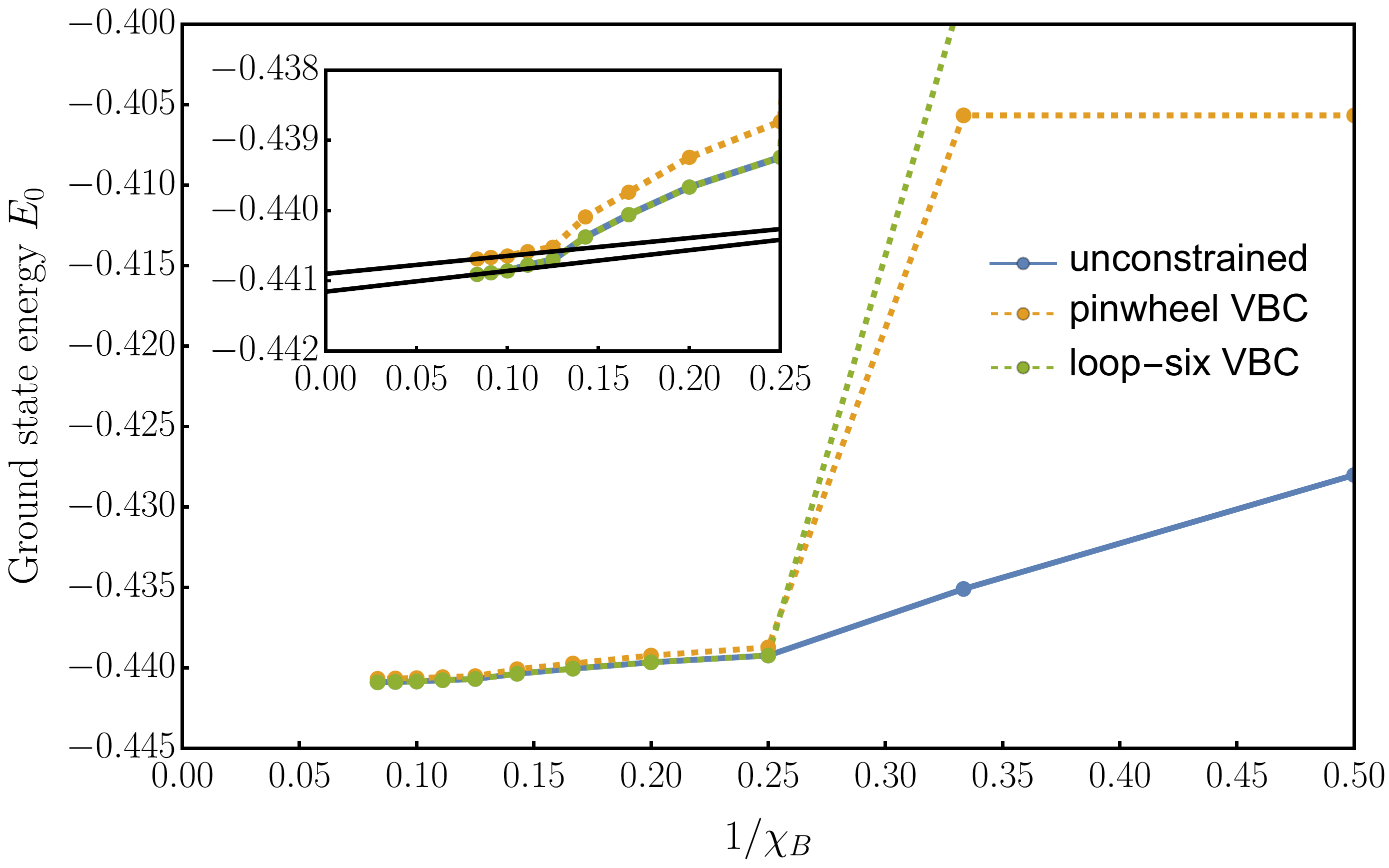}
    \caption{Comparison of unconstrained and constrained ground state simulations up to $\chi_B = 12$ for the different configurations imprinted. A first-order polynomial fit is used to extract the infinite bond dimension limit for the three largest bond dimensions.}
    \label{fig:groundStateEnergy_2}
\end{figure}

In our simulation, the main difference in the iPESS and iPEPS calculations is in the simple update. It is more local in the iPESS with only three sites that are updated at once, compared to six sites in the iPEPS ansatz [see Appendix]. This, along with a larger number of variational parameters in the iPEPS ansatz is responsible for a better ground state approximation with lower energies. For subsequent investigation of the model we therefore use the iPEPS with a $(2, 2)$ unit cell configuration. In addition to the unconstrained simulations, we incorporate fully $SU(2)$-symmetric simulations of the model. By imposing the symmetry, the simulated ground state is guaranteed to be in the spin-$0$ sector, i.e., a spin singlet. In contrast, an unconstrained simulation of the ground state can spontaneously break $SU(2)$-symmetry, which would lower its energy. An energy comparison is therefore another way to ascertain the nature of the ground state. For large enough bond dimensions, the $SU(2)$-symmetric simulations converge to the same energy as the unconstrained ones, confirming a nonmagnetic VBC ground state with spin-$0$ of the model as previously reported~\cite{Astrakhantsev2021}.

Within our iPEPS simulations, we use two ways to ascertain the nature of the VBC ground state, (i) we prepare our initial state with the pinwheel and the loop-six VBC patterns imprinted for a given low bond dimensions and progressively increase $\chi_{B}$ in a manner which uses the converged state vectors at any given $\chi_{B}$ as initial states for the simulation with one higher bond dimension and (ii) an unconstrained optimization starting from a random state. In procedure (i) we observe that while both the pinwheel VBC and loop-six VBC patterns remain stable up to $\chi_B = 12$, the latter is always lower in energy at any given finite bond dimension [see Fig.~\ref{fig:groundStateEnergy_2} and Table~\ref{tab:energyComparison_1}]. The resulting spatial spin-spin correlation profiles at $\chi_{B} = 12$ are shown in Fig.~\ref{fig:groundStateConfigurations_1}. Further compelling evidence supporting a loop-six VBC ground state scenario for a finite bond dimension is provided by the unconstrained optimization which at higher bond dimensions $(\chi_{B}\geqslant 4)$ already converges to the energy of the loop-six VBC [see Fig.~\ref{fig:groundStateEnergy_2} and Table~\ref{tab:energyComparison_1}]. Notice that the pinwheel and loop-six pattern is explicitly imprinted in the simulations at $\chi_B = [2, 3]$, so that those points are not expressive. The inset of Fig.~\ref{fig:groundStateEnergy_2} shows the meaningful, i.e., linear regime where the energy differences between the two orders are small, highlighting the subtle competition.

An extrapolation to the infinite bond dimension limit using a linear fit of the three values of energy corresponding to the largest $\chi_B$ yields a lower bound for the energy $E_l$. The last data point at $\chi_B = 12$ provides an upper bound $E_u$, such that the true ground state energy lies in the interval $\lbrack E_l, E_u \rbrack$~\cite{Corboz2DHubbard}. To estimate the final ground state energy, we compute \mbox{$E_0 = (E_u + E_l)/2$} with an error of \mbox{$\Delta E = (E_u - E_l)/2$}, which results in
\begin{align}
    \begin{split}
        E_0(\text{pinwheel VBC}) &= -0.4408 \pm 0.0001,\\
        E_0(\text{loop-six VBC}) &= -0.4410 \pm 0.0001\, ,
    \end{split}
\end{align}
which is lower than previous estimates of the ground state energy~\cite{Astrakhantsev2021}. The numerical values for the results in Fig.~\ref{fig:groundStateEnergy_2} are 
summarized in Table~\ref{tab:energyComparison_1}. Given that the estimates of the ground state energy for the pinwheel and loop-six VBC states evaluated in the limit $\chi_{B} \to \infty$ are very close, variational iPEPS simulations have been employed to resolve which of these two competing states wins in this limit. Until the largest reachable bond dimension of $\chi_B = 7$, the variational energies lie below the presented simple update energies and reinforce the ground state to be a loop-six VBC~\cite{Naumann2023}. A direct comparison of the simple update and variational energies is presented in Table~\ref{tab:energyComparison_2}.

\begin{table}[t]
    \centering
    \begin{tabular}{c c c c}
         $\chi_B$ & Unconstrained & Pinwheel & Loop-six \\
         \hline
          2 & -0.428020 & {\color{gray}-0.405664} & {\color{gray}-0.397792} \\ 
          3 & -0.435105 & {\color{gray}-0.405664} & {\color{gray}-0.397792} \\
          4 & -0.439242 & -0.438734 & -0.439242 \\
          5 & -0.439665 & -0.439242 & -0.439665 \\
          6 & -0.440058 & -0.439738 & -0.440058 \\
          7 & -0.440375 & -0.440091 & -0.440376 \\
          8 & -0.440700 & -0.440522 & -0.440700 \\
          9 & -0.440776 & -0.440584 & -0.440776 \\
         10 & -0.440859 & -0.440646 & -0.440859 \\
         11 & -0.440886 & -0.440667 & -0.440886 \\
         12 & -0.440908 & -0.440689 & -0.440908 \\
    \end{tabular}
    \caption{Energy comparison for different ground state configurations of the $(2,2)$ iPEPS. Note that the states at $\chi_B = [2, 3]$ cannot be used, since the ground state pattern is already imprinted.}
    \label{tab:energyComparison_1}
\end{table}

\begin{table}[t]
    \centering
    \begin{tabular}{c c c c}
         $\chi_B$ & Simple update & Variational update \\
         \hline
          2 & -0.428020 & -0.433600 \\ 
          3 & -0.435105 & -0.437320 \\
          4 & -0.439242 & -0.439877 \\
          5 & -0.439665 & -0.440162 \\
          6 & -0.440058 & -0.440391 \\
          7 & -0.440375 & -0.440592 \\
    \end{tabular}
    \caption{Energy comparison between simple update and variational energies for the ground state of the loop-six VBC. The variational update uses a two-tensor checkerboard pattern on the coarse-grained Shuriken lattice.}
    \label{tab:energyComparison_2}
\end{table}

\begin{figure}[b]
    \centering
    \includegraphics[width = 1.0\columnwidth]{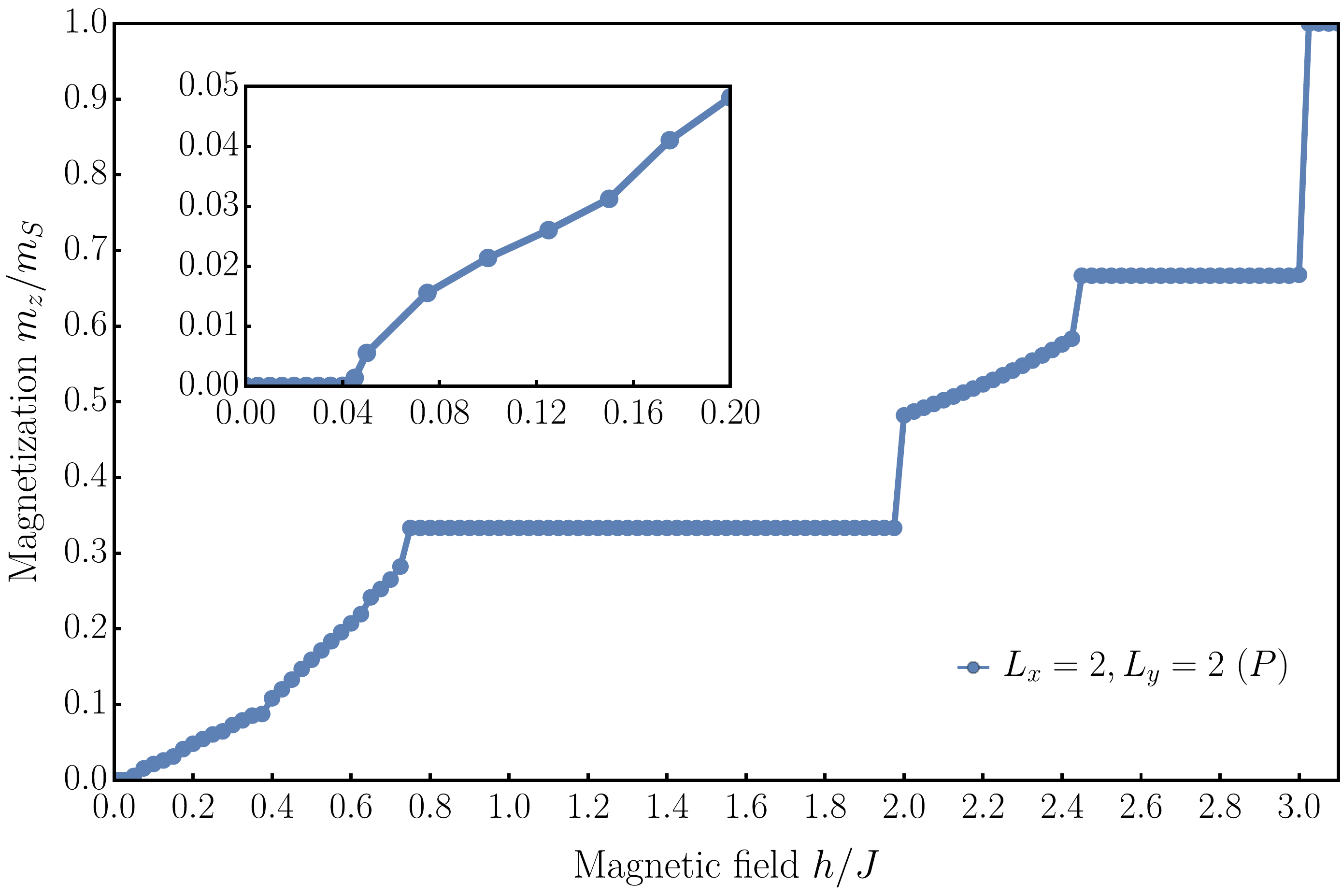}
    \caption{Magnetization curve of the Heisenberg model for $\chi_B = 10$. Upon tuning the magnetic field, two magnetization 
    plateaus at $1/3$ and $2/3$ of the saturated magnetization $m_S = 1/2$ appear. Additionally, we find the presence of a 
    small plateau at $m_z = 0$ indicative of the gapped nature of the ground state.}
    \label{fig:magnetizationPlateaus_1}
\end{figure} 

\section{Results on magnetization plateaus}
\begin{figure}[t]
    \begin{minipage}{0.85\columnwidth}
        \includegraphics[width = .85\columnwidth]{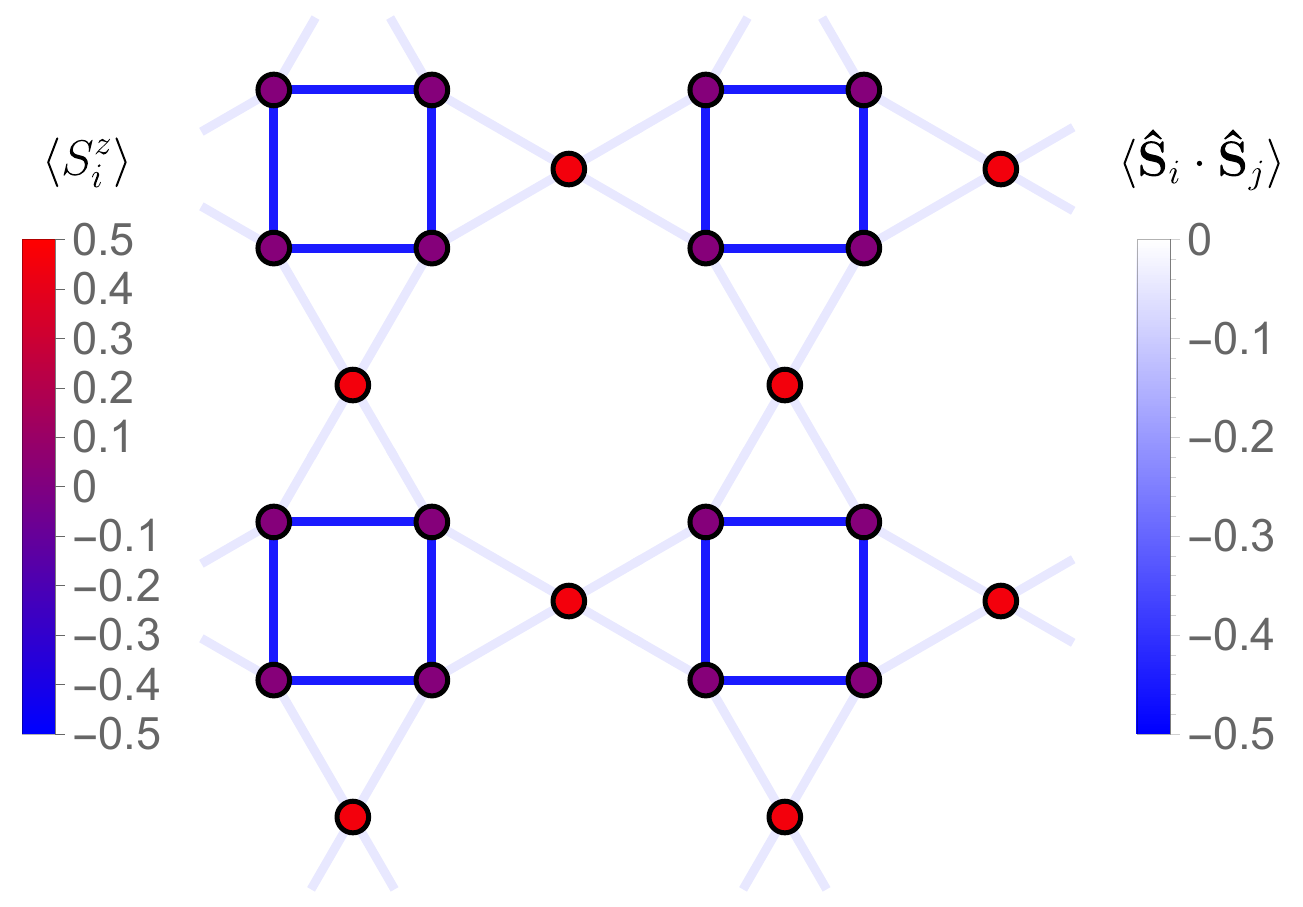}
    \end{minipage}
    \begin{minipage}{0.85\columnwidth}
        \includegraphics[width = .85\columnwidth]{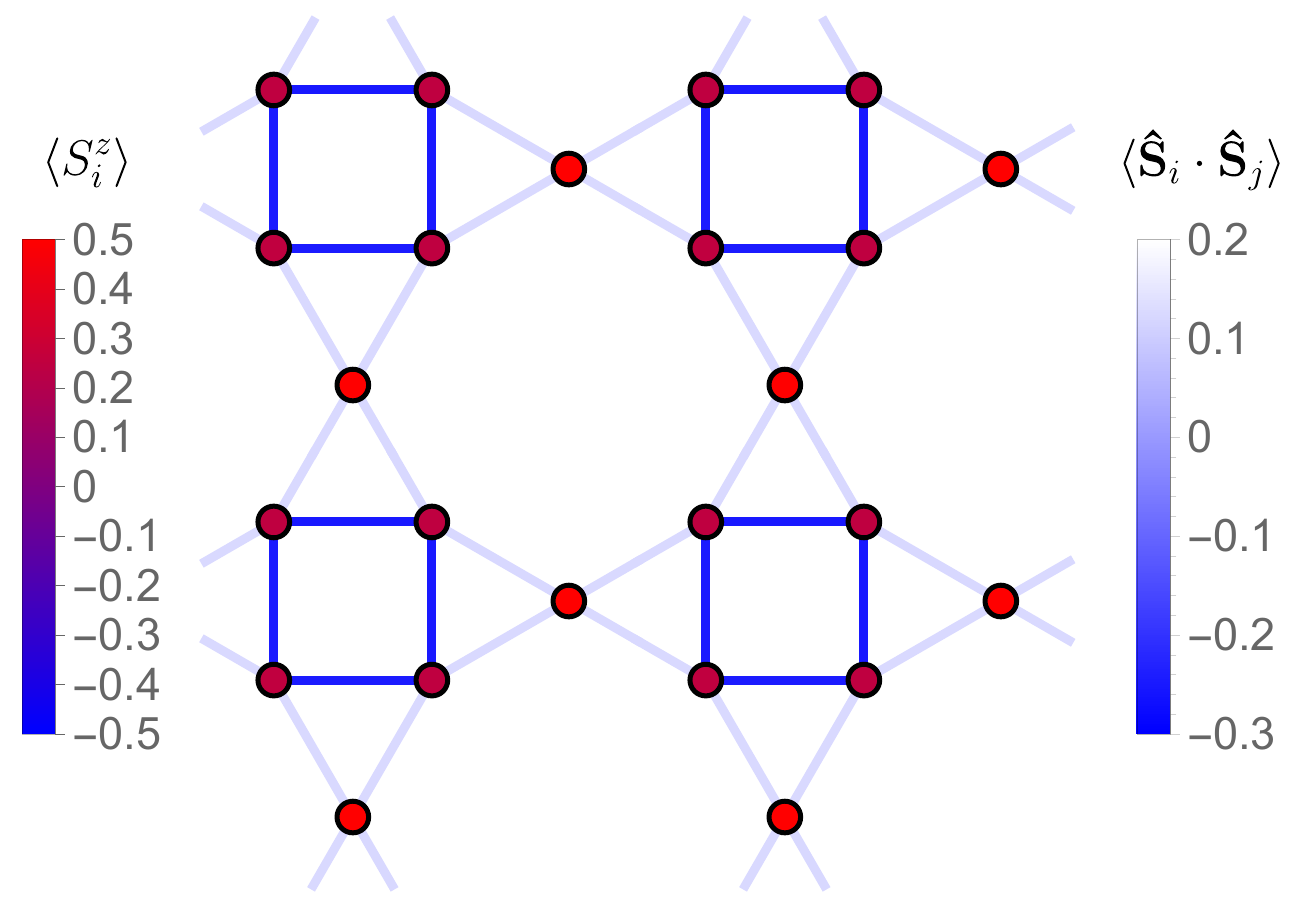}
    \end{minipage}
    \caption{Ground state configurations of magnetic plateaus for $h = 1.4$ (top) and $h = 2.8$ (bottom) at $\chi_B = 10$. Both the $1/3$ plateau state and the $2/3$ plateau state have strong spin-spin correlations on the squares, whose spins appear to be in an entangled superposition. In contrast, spins on the Shuriken sites are isolated and (almost) fully polarized.}
    \label{fig:groundStateConfigurations_2}
\end{figure}

Finally, we study the Heisenberg model on the Shuriken lattice in the presence of an external magnetic field. We compute the average magnetization over all the sites in the lattice along the field axis
\begin{align}
    m_z = \frac{1}{N} \sum_i \langle \psi_0(h) \vert \hat{S}_i^z \vert \psi_0(h) \rangle\ ,
\end{align}
where $\vert \psi_0(h) \rangle$ is the normalized ground state vector. In Fig.~\ref{fig:magnetizationPlateaus_1} we show the magnetization curve normalized to the saturation value of \mbox{$m_S = 1/2$}. The magnetization curve reveals the presence of three magnetization plateaus, at $0$, $1/3$ and $2/3$ of the saturation value~\cite{Richter-2009,Nakano-2013,Richter-2022}. Furthermore, we observe a macroscopic jump from the $2/3$ plateau to saturation magnetization as conventionally expected due to the presence of a flat one-magnon band which leads to the appearance of localized multi-magnon eigenstates~\cite{Schnack-2001,Schulenberg-2002,Derzhko-2004,Zhitomirsky-2004,Derzhko-2007,Mizoguchi-2021}. The plateau at $h \rightarrow 0$ is a further indication of the fact that the ground state of the model at $h = 0$ is actually gapped. An estimate on the size of the spin gap \mbox{$\Delta > 0$} is given by the width of the plateau \mbox{$\Delta\sim 0.04 J$}, consistent with exact diagonalization studies~\cite{Rousochatzakis-2013,Nakano-2013}. The $1/3$ and $2/3$ plateaus can further be characterized by the 
spatial pattern of spin-spin correlations and the expectation values of the spin-$x$, -$y$ and -$z$ components. Those expectation values are shown in Fig.~\ref{fig:groundStateConfigurations_2} for both phases, at magnetic fields $h = 1.4$ and $h = 2.8$ respectively. Interestingly, one observes that once the magnetic field is turned on, a pattern governed by strong loop four resonances emerges.

\begin{figure}[t]
    \centering
    \includegraphics[width = 1.0\columnwidth]{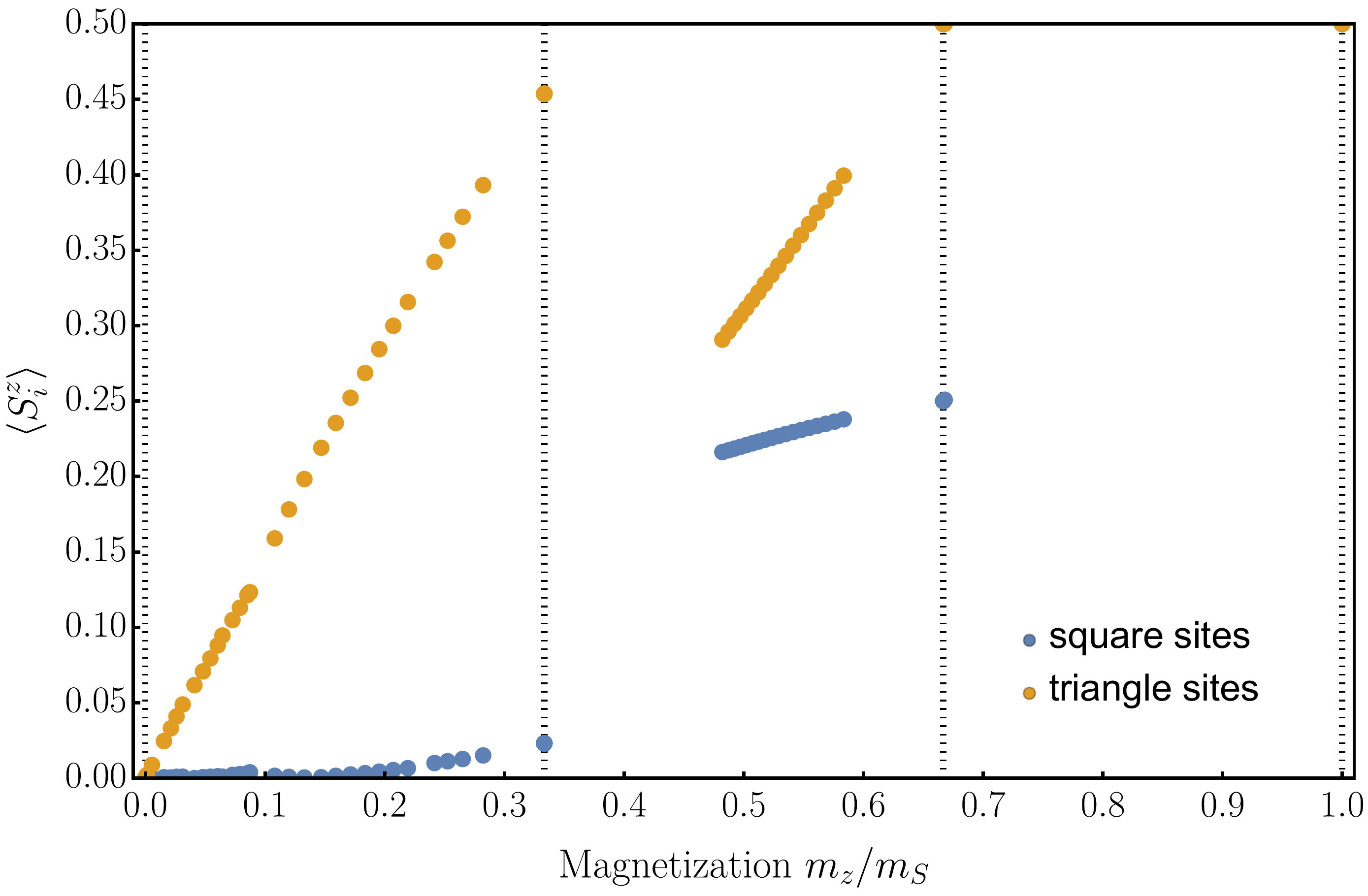}
    \caption{On-site magnetization of the triangle and square sites versus the magnetic field $h$. Dotted vertical lines represent the four magnetization plateaus. }
    \label{fig:magnetizationPlateaus_2}
\end{figure}

Within error-bars in the expectation values, the states at both magnetization plateaus are invariant under translations of the original six-site crystallographic unit cell and also under point group symmetries. It is also visible that correlations are much stronger on the squares compared to the triangle bonds in the lattice. For both plateau states the spins which are not part of the squares are isolated and almost fully polarized [see Fig.~\ref{fig:magnetizationPlateaus_2}], implying that they are nearly aligned with the magnetic field. In contrast, the spins on squares, despite a finite magnetization possess a nonzero singlet density reminiscent of $h_{z}=0$ resonating plaquettes. This is also evidenced by observing the different magnetization behaviours $\langle \hat{S}_i^z \rangle$ of the two symmetry inequivalent sites over the range of the magnetization as shown in Fig.~\ref{fig:magnetizationPlateaus_2}.
The three plateaus appearing in Fig.~\ref{fig:magnetizationPlateaus_1} and the full saturation are shown with dotted lines. The behaviour is in good agreement with previous numerical diagonalization studies of the model on finite systems~\cite{Nakano-2013}.

{\setlength{\fboxrule}{0pt}
Following the discussion of an analytic description for the magnetic plateau states of the Heisenberg model on the Kagome lattice~\cite{capponi13_numer_study_magnet_plateaus_spin}, one can similarly construct the 2/3 plateau state on the Shuriken lattice. For the 2/3 case, one can see in Fig.~\ref{fig:groundStateConfigurations_2} and \ref{fig:magnetizationPlateaus_2} that the expectation value $\langle \hat{S}_i^z \rangle$ is roughly $0.25$ for the sites which are part of the square and $0.5$ for the sites which are not part of the square. This motivates the conclusion that the state vector for one unit cell consists of an entangled state on the square and a product of this superposition with the fully polarized non-square sites as $\ket{\psi^{2/3}} = \ket{\uparrow}_{\siteOne} \ket{\framebox[0.425cm]{\raisebox{-0.15cm}{{\squareSites}}}} \ket{\uparrow}_{\siteSix}$. Using this ansatz, it is straightforward to find the ground state vector
\begin{align}
    \ket{\framebox[0.425cm]{\raisebox{-0.15cm}{{\squareSites}}}} &= \frac{1}{2} \left( \ket{\downarrow \uparrow \uparrow  \uparrow} - \ket{\uparrow \downarrow  \uparrow  \uparrow} - \ket{\uparrow \uparrow \downarrow \uparrow} + \ket{\uparrow \uparrow \uparrow \downarrow} \right)
\end{align}
for the square terms, consisting of pairwise singlets on the four bonds. Since the individual unit cells are not entangled in our ansatz it is easy to check with exact diagonalization (ED), that the full state is the ground state of the subspace we have chosen. The per-site energy of the analytic construction is $-2 h / 6$. This perfectly fits our simple update results, where for $h=2.8$ we find an energy of $-0.933331$ while the analytic result is $-0.933333$. This construction built out of one-magnon states on the squares and fully polarized spins on Shuriken sites is the well-known magnon crsytal state~\cite{Richter-2004a,Richter-2009,Schnack-2001}.}

For the 1/3 plateau the same approach is not successful. If we limit ourselves to the subspace with fully polarized spins on the non-square sites and the superposition on the square sites, we can find a ground state from ED but this time the analytic energy for $h=1.4$ of $-0.566667$ is clearly above the result $-0.592928$ of the SU calculations. This is expected however, since as one can see in Fig.~\ref{fig:magnetizationPlateaus_2} the spin expectation values $\langle \hat{S}_i^z \rangle$ are neither exactly $0.5$ for the polarized nor zero for the spins on the square. Therefore, a more sophisticated ansatz would be needed to describe this state but this beyond the scope of this work.
 
\section{Conclusions}

In this work we have studied the ground state properties of the $S=1/2$ Heisenberg anti-ferromagnet on the Shuriken lattice using two-dimensional tensor network techniques of iPEPS and iPESS. We find that the incorporation of non-local correlations proves indispensable in accurately capturing the nature of the ground state, which is shown to be a loop-six VBC at any given finite bond dimension up until $\chi_{B} = 12$. Indeed, here we are faced with a scenario featuring a delicate energetic competition between states governed by different stabilization mechanisms. In particular, we have (i) a pinwheel VBC favored by energy gain from short-range loop resonances and (ii) a loop-six VBC favored by strong resonances over longer-length loops which are amplified by the dressing of virtual singlets on top of the nearest-neighbor basis. This is precisely what we aim to address by employing the TN framework which naturally contains both these key ingredients and allows us to accurately investigate their interplay towards accurately determining the nature of the ground state. Our estimate of the ground state energy per site in the thermodynamic limit obtained by extrapolating $\chi_{B}\to\infty$ is given by $E_0 = -0.4410 \pm 0.0001$.

We have also investigated the effect of an external magnetic field in the model and obtained its magnetization curve which shows three magnetization plateaus at $0$, $1/3$ and $2/3$ of the saturation magnetization. The width of the magnetization plateau at zero-field gives us an estimate of the spin gap $\Delta \sim 0.04 J$ consistent with exact diagonalization studies~\cite{Rousochatzakis-2013,Nakano-2013}. The nature of the phases at these plateaus are not only polarized but also show strong signature of singlet correlations on four-site plaquettes. These states are found to respect the spatial symmetries (both translation and point group) of the Shuriken lattice, in contrast to the pinwheel VBC state.

Our work paves the way for future investigation of the Heisenberg model on the Shuriken lattice in more general settings. It would be interesting to study the anisotropic model with different couplings on the square and the triangle bonds, or longer range couplings which could potentially be of relevance in describing the recently studied materials~\cite{Yakubovich-2021,Fujihala-2020,Liu-2022}. An investigation of the excitation spectrum would be another promising route to revealing diverse manifestation of frustration~\cite{Clarty-2020}. Similarly, the corresponding model for higher spins, e.g., $S=1$, could be explored, which could be host to a trimerized ground state and display a wealth of magnetization plateaus hosting exotic phases. This perspective seems even more interesting as it seems plausible to recreate frustrated systems in Shuriken lattices under the precisely controlled conditions of \emph{quantum simulations}~\cite{PhysRevX.4.041037} involving ultra-cold atoms, giving rise to the interesting situation of benchmarking quantum and classical simulations against each other.

\section{Acknowledgments}
We thank Arnaud Ralko, Ronny Thomale and Erik Weerda for insightful discussions and helpful comments on the manuscript. Y.\,I. acknowledges support from the Department of Science and Technology (DST), India through the MATRICS Grant No.~MTR/2019/001042, CEFIPRA Project No. 64T3-1, the ICTP through the Associates Programme and from the Simons Foundation through grant number 284558FY19. This research was supported in part by the National Science Foundation under Grant No.~NSF~PHY-1748958, IIT Madras through the Institute of Eminence (IoE) program for establishing the QuCenDiEM group (Project No. SB20210813PHMHRD002720), the International Centre for Theoretical Sciences (ICTS), Bengaluru, India during a visit for participating in the program ``Frustrated Metals and Insulators'' (Code: ICTS/frumi2022/9). Y.~I.~acknowledges the use of the computing resources at HPCE, IIT Madras. The Berlin team has been supported by the BMBF (MUNIQC-ATOMS), the DFG (CRC 183 on ``Entangled states of matter'', project No.\ 277101999), and the Helmholtz Center Berlin. The authors would like to thank the HPC Service of ZEDAT, Freie Universität Berlin for computing time~\cite{Bennett2020}.

\appendix*
\section{Tensor networks}
\label{app:TensorNetworkDetails}

\subsection{Tensor network structures and algorithms}

\begin{figure}[ht]
    \centering
    \includegraphics[width = 0.6\columnwidth]{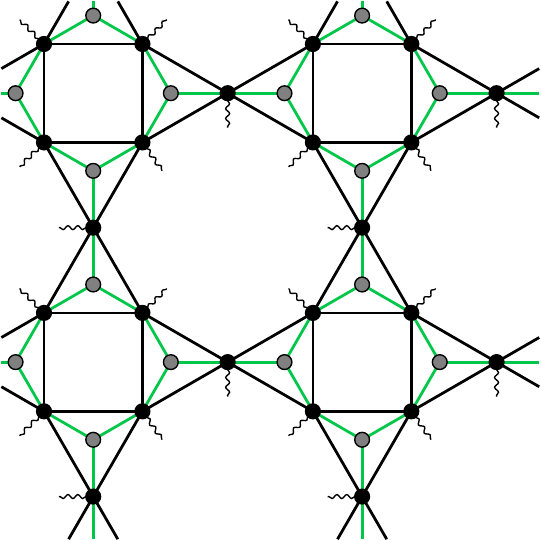}
    \caption{Shuriken lattice shown in black and its dual lattice (the so-called $(4, 8^2)$ Archimedean lattice) shown in green. Since the spin-$1/2$s live on the links of the dual lattice, additional three-index simplex tensors are introduced on the vertices for the iPESS ansatz.}
    \label{fig:SquareKagomeLattice_2}
\end{figure}

In this section, we present further technical details about the tensor network structures and algorithms employed in the main article. We will start with the \emph{infinite projected entangled simplex state} (iPESS) ansatz~\cite{Xie2014}, which can be straightforwardly extended from the original formulation on the Kagome lattice to the Shuriken lattice. To this end, we consider the dual lattice (the so-called $(4, 8^2)$ Archimedean lattice), where the spin-$1/2$s are located on the lattice links instead of the lattice sites. This lattice is visualized in Fig.~\ref{fig:SquareKagomeLattice_2} in green.

In order to connect the spins, additional purely virtual three-index simplex tensors have to be introduced (shown in gray). An elementary iPESS unit cell therefore consists of six lattice site tensors carrying the physical degree of freedom, and four simplex tensors connecting them.

\begin{figure}[ht]
    \centering
    \includegraphics[width = 0.9\columnwidth]{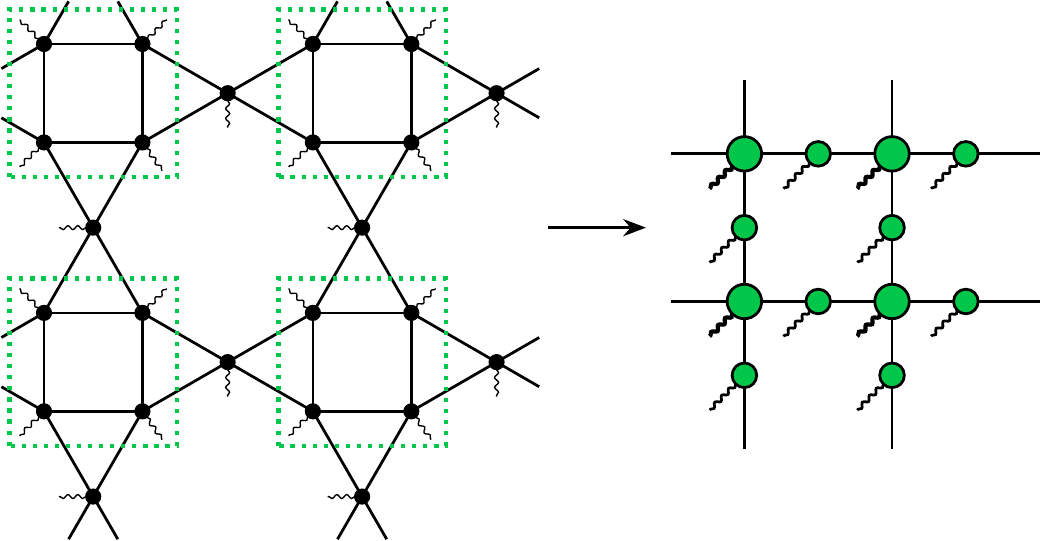}
    \caption{Shuriken lattice shown on the left and iPEPS ansatz shown on the right. The four sites on each square are coarse-grained into an effective site, highlighted by the green dotted area. The resulting structure is a square lattice with missing links.}
    \label{fig:SquareKagomeLattice_1}
\end{figure}

As an alternative approach we consider a modified version of the \emph{infinite projected entangled pair state} (iPEPS). It is constructed by a partial coarse-graining of the original lattice to a square lattice with missing bonds. To this end we merge the four spins on each square configuration into a single site, which is modeled by a tensor with physical dimension $p^4$. The two remaining sites per unit cell are left unchanged. This mapping is shown in Fig.~\ref{fig:SquareKagomeLattice_1}.

One advantage that both TN structures share is the fact that one virtual bond in the TN corresponds to only two links in the original Shuriken lattice. Since the maximal entanglement shared between neighbouring lattice sites is limited by the bond dimension of the TN ansatz, it is favourable to keep this number small. Due to this property, we can directly compare results with the same bond dimension. Note that a coarse-graining of the six spins per unit cell in the Shuriken lattice directly results in a regular square lattice.

\subsection{Simple update}

\begin{figure}[htb]
    \centering
    \includegraphics[width = 1.0\columnwidth]{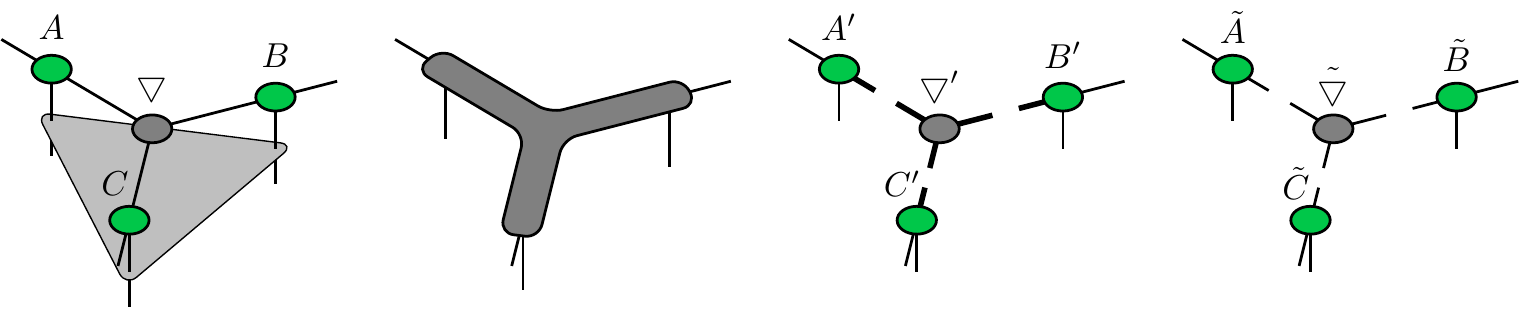}
    \caption{Simple update in the iPESS simulations of the Shuriken lattice. The Trotterized three-body Hamiltonian gate is absorbed into a triangle configuration. A truncated higher-order SVD is used to decompose the resulting six-index tensor back into the separate iPESS tensors. The same procedure is applied to the other simplex tensors along with the different lattice tensors.}
    
    \label{fig:simpleUpdate_SquareKagome_1}
\end{figure}

In order to obtain an approximation of the ground state wave function, we employ the simple update technique~\cite{simpleupdatejiang} in both TNs. This method is based on an imaginary-time evolution under the Hamiltonian~\cite{Jordan2008}, in which all the tensors in the networks are updated sequentially. For a sufficiently long evolution, the ground state is projected out. For the iPESS network we employ a regular three-site update of the different simplex configurations~\cite{Xie2014}. In order to restore the individual tensors after each update, a higher-order singular value decomposition (SVD) is used. During this process, the singular values are truncated to the fixed bulk bond dimension $\chi_B$ to keep the simulations computationally feasible. The process is illustrated in Fig.~\ref{fig:simpleUpdate_SquareKagome_1} for the update of a down simplex, denoted $\bigtriangledown$, together with the three connected lattice tensors.

\begin{figure}[htb]
    \centering
    \includegraphics[width = 1.0\columnwidth]{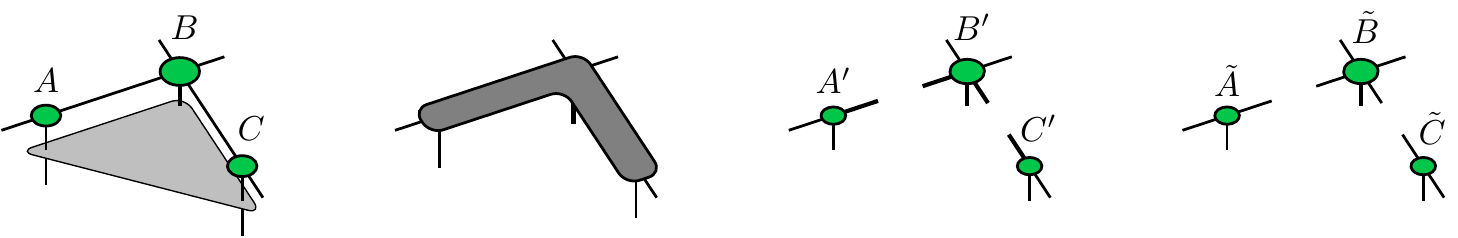}
    \caption{Simple update in the iPEPS simulations of the Shuriken lattice. The Trotterized six-body Hamiltonian gate is absorbed into a bottom-left corner of the deformed square lattice. The individual tensors are restored using two successive SVDs with truncation to a fixed bond dimension. The update of a top-right corner is performed similarly.}
    \label{fig:simpleUpdate_SquareKagome_2}
\end{figure}

The simple update for the iPEPS ansatz on the deformed square lattice follows the same spirit. However, instead of only updating three sites (along with one simplex tensor) as in the iPESS ansatz, we choose a six-site update across corners in the lattice. A single update step is presented in Fig.~\ref{fig:simpleUpdate_SquareKagome_2}.

Again, the virtual links of the network are kept at a maximal bond dimension $\chi_B$, which is achieved by two successive SVDs. Notice that we omit to show additional diagonal tensors carrying the singular values on each virtual link for more clarity both in Fig.~\ref{fig:simpleUpdate_SquareKagome_1} and Fig.~\ref{fig:simpleUpdate_SquareKagome_2}.\\

Besides the two presented TN approaches, we also implemented a simple update scheme on the original Shuriken lattice, using two-body gates on neighbouring sites to evolve the wave function. Similarly as for the iPESS, the update is very local and could not resolve all the magnetization plateaus present in the model, so that we rejected the simulations.

\subsection{Environments and expectation values}

In order to compute accurate expectation values for the wave functions obtained by the simple update, we employ a \emph{corner transfer matrix renormalization group} (CTMRG)~\cite{ctmnishino1996,ctmnishino1997,ctmroman2009,ctmroman2012} procedure to compute the effective environments. To this end, the six lattice sites per unit cell are coarse-grained into a single iPEPS site with local Hilbert space dimension $2^6 = 64$. This maps the Shuriken lattice to a regular square lattice, for which a directional CTMRG procedure can directly compute the approximate contraction of the infinite lattice.

\begin{figure}[ht]
    \centering
    \includegraphics[width = 1.0\columnwidth]{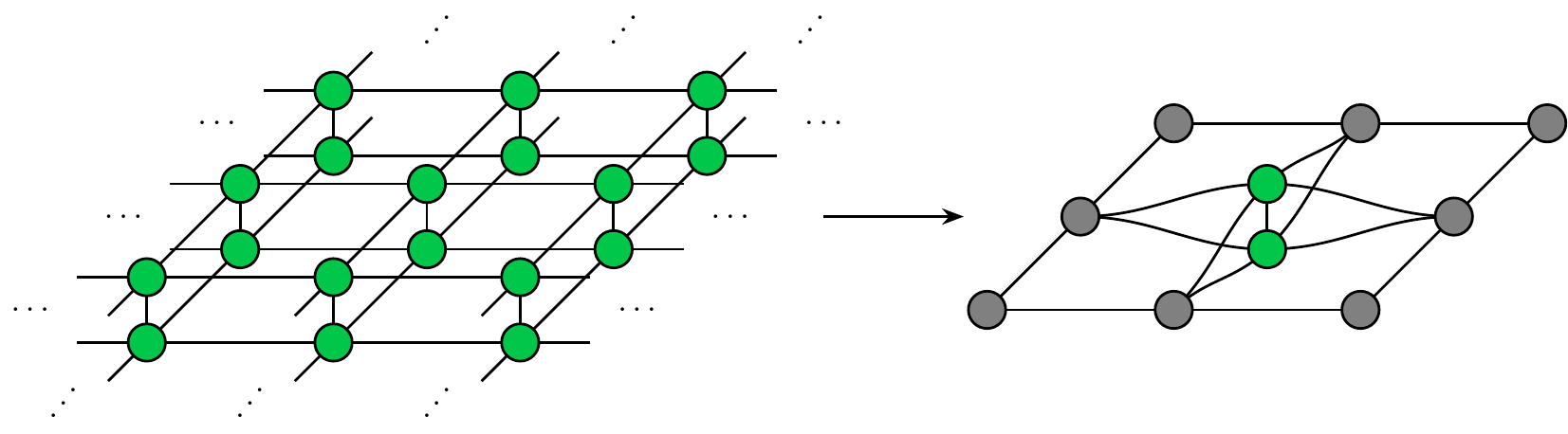}
    \caption{A CTMRG routine is used to approximate the contraction of the infinite square lattice by a set of fixed-point environment tensors.}
    \label{fig:CTMRG_Method_1}
\end{figure}

As shown in Fig.~\ref{fig:CTMRG_Method_1}, the contraction of the infinite square lattice is approximated by a set of eight fixed-point tensors surrounding every iPEPS tensor in the unit cell. Expectation values can then be computed straightforwardly by evaluating local operators $\langle \psi \vert \hat O \vert \psi \rangle / \langle \psi \vert \psi \rangle$, where the environment around the sites on which the operator acts and the norm of the wave function is approximated by the CTMRG tensors.

\newpage

%

\end{document}